\documentclass{emulateapj}

\usepackage{natbib}
\usepackage{amssymb}
\def\arctanh{\mathrel{\rm arctanh}}

\shorttitle{Stellar and total baryon mass fraction in groups and clusters since z=1}
\shortauthors{Giodini et al.}

\slugcomment{Submitted for publication in the Astrophysical Journal}

\begin{document}

\title{Stellar and total baryon mass fractions in groups and clusters since redshift 1}\altaffiltext{$\star$}{Based on observations
   obtained with XMM-Newton, an ESA science mission with
instruments and contributions directly funded by ESA Member States and NASA;
also based on data collected at: the NASA/ESA {\em
Hubble Space Telescope}, obtained at the Space Telescope Science
Institute, which is operated by AURA Inc, under NASA contract NAS
5-26555; the Subaru Telescope, which is operated by
   the National Astronomical
Observatory of Japan; the European Southern Observatory, Chile, under Large
   Program 175.A-0839; and the Canada-France-Hawaii Telescope
   operated by the National Research Council of Canada,
the Centre National de la Recherche Scientifique de France and the University
   of Hawaii.}

\author{S.Giodini \altaffilmark{1}, D. Pierini \altaffilmark{1},
A. Finoguenov \altaffilmark{1,2},
G. W. Pratt \altaffilmark{1},
H. Boehringer \altaffilmark{1},
A. Leauthaud\altaffilmark{7},
L. Guzzo \altaffilmark{3},
H. Aussel \altaffilmark{13},
M. Bolzonella \altaffilmark{18},
P. Capak \altaffilmark{4,14},
M. Elvis \altaffilmark{5},
G. Hasinger\altaffilmark{17},
O. Ilbert\altaffilmark{6},
J. S. Kartaltepe\altaffilmark{6},
A. M. Koekemoer\altaffilmark{16},
S. J. Lilly \altaffilmark{8},
R. Massey \altaffilmark{14},
H. J. McCracken \altaffilmark{9},
J. Rhodes \altaffilmark{14,19},
M. Salvato \altaffilmark{14},
D. B. Sanders\altaffilmark{6},
N. Z. Scoville\altaffilmark{4},
S. Sasaki\altaffilmark{10,11},
V. Smolcic\altaffilmark{14},
Y. Taniguchi\altaffilmark{12},
D. Thompson\altaffilmark{14, 15}
and the COSMOS collaboration
}
\altaffiltext{1}{Max Planck Institut f{\"u}r Extraterrestrische Physik, Giessenbachstrasse, Garching bei M{\"u}nchen D-85748, Germany}
\altaffiltext{2}{University of Maryland, Baltimore County, 1000 Hilltop Circle, Baltimore, MD 21250.}
\altaffiltext{3}{INAF-Osservatorio Astronomico di Brera, Via Bianchi 46, I-23807 Merate (LC), Italy}
\altaffiltext{4}{Spitzer Science Center, 314-6 Caltech, Pasadena, CA 91125}
\altaffiltext{5}{Harvard-Smithsonian Ctr. for Astrophysics (USA)}
\altaffiltext{6}{Institute for Astronomy, University of Hawaii, 2680 Woodlawn Drive, Honolulu, HI 96822,USA}
\altaffiltext{7}{LBNL \& Berkeley Center for Cosmological Physics,
 University of California, Berkeley, CA 94720, USA}
\altaffiltext{8}{Institute of Astronomy, Department of Physics, Eidgenšssische Technische Hochschule, ETH Zurich, CH-8093, Switzerland}
\altaffiltext{9}{Institut d'Astrophysique de Paris, UMR 7095 CNRS, UniversitŽ Pierre et Marie Curie, 98bis boulevard Arago, 75014 Paris, France}
\altaffiltext{10}{Astronomical Institute, Graduate School of Science, Tohoku University, Aramaki, Aoba, Sendai 980-8578, Japan.}
\altaffiltext{11}{Physics Department, Graduate School of Science and Engineering, Ehime University, 2--5 Bunkyo--cho, Matsuyama 790--8577, Japan.}
\altaffiltext{12}{Research Center for Space and Cosmic Evolution, Ehime University, 2-5 Bunkyo-cho, Matsuyama 790-8577, Japan}
\altaffiltext{13}{ AIM Unit\'e Mixte de Recherche CEA  CNRS Universit\'e Paris VII UMR n158}
\altaffiltext{14}{California Institute of Technology, MC 105-24, 1200 East California Boulevard, Pasadena, CA 91125}
\altaffiltext{15}{Large Binocular Telescope Observatory, University of Arizona, 
Tucson, AZ85721, USA }
\altaffiltext{16}{Space Telescope Science Institute, 3700 San Martin Drive, Baltimore, MD 21218}
\altaffiltext{17}{Max-Planck-Institut f{\"u}r Plasmaphysik, Boltzmannstrasse 2, Garching bei M{\"u}nchen D-85748, Germany }
\altaffiltext{18}{INAF - Bologna Astronomical Observatory, via Ranzani 1, I-40127 Bologna, Italy}
\altaffiltext{19}{Jet Propulsion Laboratory, California Institute of Technology, Pasadena, CA 91109, USA}

\begin{abstract}
We investigate if the discrepancy between estimates of the total baryon
mass fraction obtained from observations of the cosmic microwave
background (CMB) and of galaxy groups/clusters persists when a large
 sample of groups is considered. 
 To this purpose, 91 candidate
X-ray groups/poor clusters at redshift $0.1 \le z \le 1$ are selected from
the COSMOS 2 deg$^2$ survey, based only on their X--ray luminosity and extent. This sample is complemented by 27
nearby clusters with a robust, analogous determination of the total and stellar mass inside R$_\mathrm{500}$. The total sample of 118 groups and clusters with
$z \le 1$ spans a range in M$_\mathrm{500}$ of $\sim10^{13}$--$10^{15}~\mathrm{M}_{\odot}$. We find that the stellar mass
fraction  associated with galaxies at R$_\mathrm{500}$ decreases with
increasing total mass as M$_\mathrm{500}^{-0.37 \pm 0.04}$,
independent of redshift. Estimating the total gas mass fraction from a recently derived, high quality scaling relation, the total baryon mass fraction ($f_\mathrm{500}^\mathrm{stars+gas}=f_\mathrm{500}^\mathrm{stars}+f_\mathrm{500}^\mathrm{gas}$)
is found to increase by $\sim 25$\% when M$_\mathrm{500}$ increases from
$\langle$M$\rangle=5 \times 10^{13}~\mathrm{M}_{\odot}$ to $\langle$M$\rangle = 7 \times 10^{14}~\mathrm{M}_{\odot}$. After consideration of a plausible contribution due to intra--cluster light (11--22$\%$ of the total stellar mass), and gas depletion through the hierarchical assembly process (10$\%$ of the gas mass), the estimated values of the total baryon mass fraction are still lower than the latest CMB measure of the same quantity (WMAP5), at a significance level of 3.3$\sigma$ for groups of $\langle$M$\rangle=5 \times 10^{13}~\mathrm{M}_{\odot}$. The discrepancy decreases towards higher total masses, such that it is 1$\sigma$ at $\langle$M$\rangle= 7 \times 10^{14}~\mathrm{M}_{\odot}$. We discuss this result in terms of non--gravitational processes such as feedback and filamentary heating. 

\end{abstract}



\section{Introduction}

The baryon mass fraction is a parameter which can be constrained by the primordial light element abundance set  by the nucleosynthesis at very early times. 
It has been measured to a very high precision from the 5 years \textit{Wilkinson Microwave Anisotropy Probe} (WMAP5) observations of the Cosmic Microwave Background (CMB), giving a value of $f_\mathrm{b}^\mathrm{WMAP5} = 0.171 \pm 0.009$ \citep{dunkley08}\footnote{When the WMAP5 data are combined with the distance measurements from the Type Ia supernovae (SN) and the Baryon Acoustic Oscillations (BAO), $f_\mathrm{b} = \Omega_\mathrm{b}/\Omega_\mathrm{m} = 0.1654 \pm 0.0062$ \citep{komatsu08}.}.  An independent measure of this quantity can also be achieved with galaxy clusters. These structures are large enough to be representative of the baryon content of the universe, which exists mainly in the form of X--ray emitting gas and stars. In the absence of dissipation, they are expected to provide a baryon mass fraction f$_{b}$ comparable to the one measured from the CMB \citep{white93, evrard97}.

Galaxy systems appear in a wide range of masses, from $\sim10^{13}$ to $\sim10^{15}~$M$_{\odot}$.  In a hierarchical scenario \citep{white91} the less massive ones, (M$<10^{14}$M$_{\odot}$, referred as groups) are the building blocks for the most massive ones (clusters). 
However, the vast majority of the attempts to estimate the  baryon mass fraction in nearby clusters have reported smaller values than expected \citep{ettori03, lin03, biviano06, mccarthy07}.  In addition this discrepancy appears to be larger for groups than for  clusters \citep{lin03}.
Explanations for  this invoke
physical processes which lower $f_\mathrm{b}$ in clusters
relative to the universal fraction \citep[see e.g.][]{bialek01, he06},
baryon components that fail detection by standard X-ray
and/or optical techniques \citep[see][]{ettori03, lin04},
or a systematic underestimate of $\Omega_\mathrm{m}$ by WMAP
\citep{mccarthy07}.

\citet{mccarthy07} extensively discuss possible explanations for the missing baryons. They concluded that the observed stellar mass function limits the contribution by low mass stars and brown dwarfs to a negligible fraction of the total stellar mass; furthermore they rule out a contribution by large  amounts of centrally concentrated gas, on the bases of inconsistencies with current X--ray data and the assumption of hydrostatic equilibrium.
 Consideration of the so called intra--cluster light (ICL) results into a discrepancy at the 3.2$\sigma$ level with respect to WMAP3 across the mass range 6$\times$10$^{13}$ --10$^{15}$M$_{\odot}$ \citep{gonzalez07}. As discussed by these authors, systematics may help reconciling their results with the WMAP estimate.

 In this respect, the correct determination of the gas mass fraction may be crucial. In fact, studies of the individual baryon components (stars associated with galaxies and gas)  have shown that the stellar ($f_\mathrm{500}^\mathrm{stars}=$M$_\mathrm{500}^\mathrm{stars}/$M$_\mathrm{500}$) and gas mass fractions within R$_\mathrm{500}$ \footnote{R$_\mathrm{\Delta}$ ($\Delta$=500,200,2500) is the radius within which the mass density
of a group/cluster is equal to $\Delta$ times the critical density ($\rho_{c}$)
of the Universe. Correspondingly, M$_\mathrm{\Delta}=\Delta\,\rho_{c}(z)\,(4\,\pi/3)R_{\Delta}^3$ is the mass inside R$_\mathrm{\Delta}$.} ($f_\mathrm{500}^\mathrm{gas}=$M$_\mathrm{500}^\mathrm{gas}$/M$_\mathrm{500}$) exhibit opposite behaviours as a function of the total system mass. In particular clusters have a higher gas mass fraction than groups (\citealt{vikhlinin06}; \citealt{arnaud07}; \citealt{sun08}), but a lower stellar mass fraction \citep{lin03}. This has been interpreted as a difference in the star formation efficiency between groups and clusters (\citealt{david90}; \citealt{lin03}; \citealt{lagana08})

On the other hand the mass dependence of the gas fraction and the discrepancy between the baryon mass fraction in groups/clusters and the WMAP value can be understood in terms of non--gravitational processes.
In fact AGN--heating (which can drive the gas outside the potential well) or gas pre--heating  (which inhibites the gas from falling towards the center of the potential) can explain the lack of gas within r$_{500}$ in groups. 
Therefore groups appear as the critical systems to assess the universality of the baryon fraction, and to understand complex physical processes affecting both the gas and the stellar components. 

	Little work has been conducted on estimation of the baryon mass fraction at the group regime, mainly because of  the lack of groups in existing catalogues and the difficulty of estimating masses for the individual components and the total. An insufficient sampling of the range in total mass spanned by groups and clusters is problematic for studying their overall properties in terms of mean and scatter of the population \footnote{The conclusions of \citet{lin03} and \citet{gonzalez07}, for example, are based only on, respectively, 27 and 23 systems, but only 3 and 5 of them are less massive than $10^{14}~h^{-1}~\mathrm{M}_{\odot}$.}.
A galaxy group/cluster is the result of the assembly history of the dark matter halo, as well as of the star formation processes affecting the gas. Both processes lead to multivariate outcomes and produce a large intrinsic scatter in the distribution of the observed properties of groups and clusters. 
Therefore it is essential to have a large enough sample to be representative of the population, and unbiased by selection effects, to be able to investigate the mean trend precisely.

Once such a sample is available, interesting questions to address are: (1) How does the stellar mass fraction behave across the total range of masses? (2) Does the relation between the stellar mass fraction and the total system mass evolve with redshift? (3) How does the gas mass fraction change as a function of the system total mass? (4) Is the total baryonic fraction in groups/clusters of galaxies consistent with the WMAP5 value?

In this paper we select the currently largest X-ray selected sample of groups from the COSMOS 2 deg$^{2}$ survey which consists of 91 high--quality systems at $0.1 \le z \le 1$. 
Existing observations currently do not  give constraints on the evolution of the baryonic components in individual systems at z$\ge$0.1. Our data allow us to put constraints on the redshift evolution of the average stellar fraction with mass, which we find to be consistent with zero ($\S$4.2).
Observational constraints on the evolution of the average gas mass fraction also suggest zero evolution in the cluster regime \citep{allen04}. We assume that this is applicable to our groups in the absence of observations to the contrary and we note that simulations support this hypothesis \citep{kravtsov05}.

We complement our sample with 27 nearby clusters investigated by \cite{lin03} in order to achieve a span of two orders of magnitude in total mass ($10^{13}<$M$<10^{15}~$M$_{\odot}$). In $\S$\ref{analysis} the total mass of stars associated with galaxies is directly determined for each group, and we investigate the relation between the stellar mass fraction and the total mass of the system. In $\S$4 we combine the stellar mass fraction estimates with the most recent determination of the relation between gas mass fraction and total mass based on a compilation of 41 local ($z\le$0.2) X--ray groups and clusters,
spanning the same range in mass as ours \citep{pratt08}, and we compute the total baryon fraction. 
We discuss results in $\S$\ref{discussion}.

We adopt a $\mathrm{\Lambda}$CDM cosmological model
($\Omega_{\mathrm{m}} = 0.258$, $\Omega_{\mathrm{\Lambda}} = 0.742$)
with H$_\mathrm{0} = 72~\mathrm{km~s^{-1}~\mathrm{Mpc}}^{-1}$, consistently with WMAP5 \citep{dunkley08, komatsu08}. Unless otherwise stated all quantities are estimated at an overdensity of 500.

\section{The sample}\label{sample}

\subsection{The COSMOS survey of groups/poor clusters}
The {\it Cosmic Evolution Survey} (COSMOS, \citealt{scoville07a}) was designed to probe
how galaxies, active galactic nuclei (AGN),
and dark matter evolve together within the large-scale structure.
The survey is based on multi-wavelength imaging and spectroscopy
from X-ray to radio wavelengths and covers a 2 deg$^{2}$ area,
including HST imaging of the entire field \citep{koekemoer07}. Large-scale structures in the COSMOS field have been characterized in terms of galaxy overdensity using photometric redshifts \citep{scoville07b}, weak lensing convergence maps \citep{massey07}, diffuse X-ray emission \citep{finoguenov07} and a combination of these \citep{guzzo07}.
In particular, the entire COSMOS region was imaged
through 54 overlapping XMM-{\em Newton} pointings (1.5 Ms, \citealt{hasinger07}).
Additional {\em Chandra} observations (1.8 Ms, \citealt{Elvis06})
mapped the central region to higher resolution.

In this study we use X--ray detection, gravitational lensing signal, optical photometric and spectroscopic data of the clusters and groups identified in the COSMOS survey.
The X-ray data reduction is described in detail in \citet{finoguenov07} and Finoguenov et al. (in preparation).
From a composite mosaic of the XMM-{\em Newton} and {\em Chandra} X-ray data,
it has been possible to detect and measure the flux of extended sources
(i.e., groups and clusters) down to a limit
of 10$^{-15}~\mathrm{erg~s^{-1}~cm^{-2}}$, as described
in the corresponding catalogue (Finoguenov et al. in preparation).
Extended source detection was based
on a wavelet scale-wise reconstruction of the image, as
described in Vikhlinin et al. (1998b), employing angular scales
from $8^{\prime \prime}$ to $2.1^{\prime}$.
Clusters and groups of galaxies were effectively selected by the spatial extent
of their X-ray emission, following the approach of Rosati et al. (1998),
Vikhlinin et al. (1998b), and Moretti et al. (2004).
The cluster detection algorithm consists of:
(1) removal of the background, (2) detection of AGN, (3) removal of AGN flux from large scale, and (4) search for extended emission.
As a result, a total of 219 X-ray extended sources were identified
in the redshift range $0 < z < 1.6$;
they span the rest--frame 0.1--2.4 keV  luminosity range
$10^{41} \le L_\mathrm{X} \le 10^{44}~\mathrm{erg~s^{-1}}$, which is typically  populated by groups and poor clusters.

Quality flags tag individual systems. Flag 1 is assigned to objects whose center corresponds to the X--
ray 
peak of the source, while flag 2 objects have their spectral extraction region redefined to include only their robust association with a unique optical system. 
A redshift was assigned to each candidate X--ray group/cluster, 
corresponding to the mean of the photometric redshift (photo-$z$) distribution
of the red--sequence galaxies as identified in Tanaka et al. (in preparation), if present, 
and lying within the X--ray overdensity contour region. This redshift is checked against the available spectroscopic redshifts mostly provided by the $z$COSMOS spectroscopic survey \citep{lilly07}. The presence of a red--sequence is not required for the group/cluster detection: if no overdensity of red sequence galaxies is found in the photo--$z$ space, the spectroscopic data only are checked for the presence of a galaxy overdensity in the same area. Flag 3 is assigned to high--z (z$>$1) not spectroscopically confirmed candidate groups. Flag 4 is assigned when multiple optical counterparts are present within the X--ray overdensity contour region. In this study  only systems with 
quality flag 1 or 2 are considered. 

The galaxy--group detection is irrespective of any optical characteristic, being based only on the presence of an X-ray extended source. The X-ray selection is an approximate selection by halo mass,  due to the tight X-ray luminosity--mass relation \citep{pratt08}; in this regard our selection is thus unbiased with respect to both the optical properties of the groups in our sample and the X--ray characteristic of the systems.\\
The purposes of the present study lead us to introduce three further selection criteria:
(1) only candidate groups/clusters detected in X--rays
with a significance higher than $3 \sigma$ on the flux determination are considered.
Selection of the most robust candidates
minimizes contamination by loose galaxy aggregations
or superposition of AGN along the line of sight.
(2) Only X-ray extended sources
with $L_\mathrm{X} > 10^{42}~\mathrm{erg~s^{-1}}$ are considered,
in order to limit contamination from starburst galaxies \citep{grimm03}
or field elliptical galaxies with X-ray halos \citep{diehl07}.
(3) We limit the redshift range to $0.1 \le z \le 1.0$,
where photo-$z$ of individual galaxies are most robust \citep{ilbert08}; furthermore, in this range the quality of the photo--z is equivalent to that of low resolution spectroscopy.

\begin{figure}[!h]
{\centering
\includegraphics[width=\columnwidth]{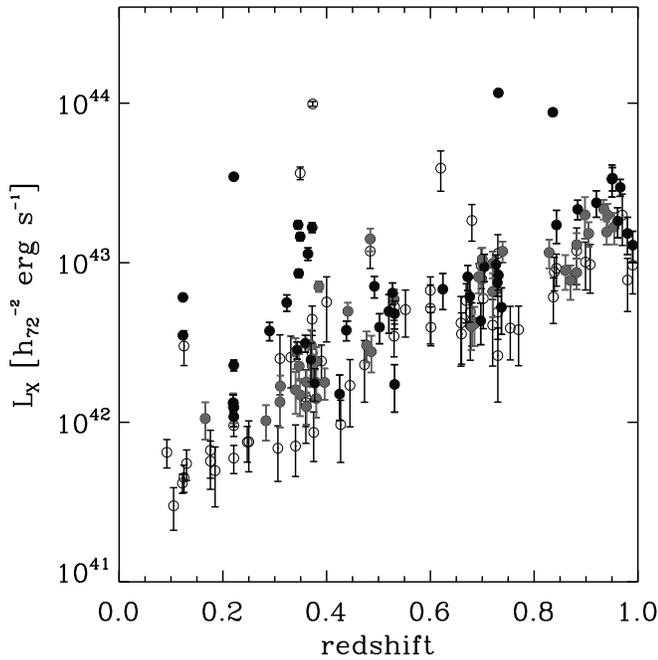}
\caption{Rest--frame 0.1--2.4 keV luminosity vs. redshift for the 151 COSMOS candidate X--ray groups/clusters at 0.1$<z<$1.0. Filled circles mark the 91 objects considered in this study: dark or light grey identifies objects with flag 1 (45) or 2 (46), respectively.\label{fig1}}}
\end{figure}

Figure \ref{fig1} reproduces the X-ray luminosity distribution
as a function of redshift for the candidate X-ray groups/clusters within z$=1$ (151 out of 219 systems).
The flag 1+2 sample selected for this study contains 114 objects, of which
44 were present in \citet{finoguenov07}.
It contains only 3 systems at $z \le 0.2$ (Figure \ref{fig1}),
which is the redshift range covered by analogous studies
on $f_\mathrm{b}$ in groups/clusters \citep{lin03, gonzalez07}. 
On the other hand, it contains systems with particularly low X-ray luminosities
(i.e., with $10^{42} < L_\mathrm{X} < 5 \times 10^{42}~\mathrm{erg~s^{-1}}$),
though only for $z < 0.5$. The sample considered in this study is reduced to 91 objects after 
removal of 23 groups with unreliable estimates of the total stellar mass in galaxies ($\S$ \ref{stellarfraction}).
Out of these 91 candidate groups/poor clusters, 51 are already spectroscopically confirmed (i.e. are associated with at least 3 galaxies with similar spectroscopic redshifts).

\subsection{COSMOS X-ray-selected groups/poor clusters: total mass estimate}\label{masses}

In the original catalogue (Finoguenov et al. in preparation),  M$_\mathrm{200}$  is computed using an L$_\mathrm{X}$--M$_\mathrm{200}$ relation established 
via the weak lensing analysis in
Leauthaud et al. (in preparation). Briefly, the COSMOS group sample is divided
into nine bins that span the redshift range $0.1<$z$<0.9$ and with
$10^{41.8}~ <$L$_\mathrm{X}$/E(z)$<10^{43.5}~$erg s$^{-1}$, where the function E(z)$\equiv
\sqrt[2]{\Omega_\mathrm{m}\mathrm(1+z)^3+\Omega_\mathrm{\Lambda}}$ represents the Hubble
parameter evolution for a flat metric. Only systems with a clear, visually identified BCG are used for this analysis, to minimize issues due to incorrect centering. 
For each bin, the weak lensing signal is calculated from $r\sim 50$ kpc to $r\sim 3$ Mpc in logarithmically spaced radial bins. A weak lensing signal is detected all the way to $3$ Mpc ensuring that the lens density is probed well beyond the virial radius. The results are fit with a parametric model which is the sum of a NFW profile \citep{navarro97} and a point--source term due to the mass of the central BCG. The theoretical relation between mass and concentration from \citet{zhao08} has been used in the fit for the NFW component and the mean stellar mass of the central BCG's is used in order to scale the point source term. A comparison between the relation obtained from the combination of the the COSMOS data and cluster data from Hoekstra et al. 2007 is consistent with that obtained by \citet{rykoff08} based on SDSS data. We adopt the following functional form for the $L_\mathrm{X}$--M relation,
\begin{equation}
 \frac{M_{200}~E(z)}{M_0} = A \left( \frac{L_{X}~E(z)^{-1}}{L_{X,0}} \right)^{\alpha}
\end{equation}
\noindent where $M_0 =10^{13.7}~$M$_{\odot}$, $L_{X,0}=10^{42.7}~$erg s$^{-1}$. Fitting only the COSMOS data yields the best fit parameters $\log_{10}(A)=0.106 \pm 0.053$ and $\alpha=0.56 \pm 0.12$ (cited errors are statistical only).
Further details regarding
the weak lensing analysis in COSMOS can be found in Leauthaud et al. in preparation.

The baryon fraction in groups/clusters can be studied at any radius, though it is desirable to study it at the largest radius possible with respect to the virial radius of the system because of the radial dependencies of the different components. The largest radius for which reliable X-ray hydrostatic masses are available is R$_{500}$ \citep[e.g.][]{arnaud05,vikhlinin06, sun08}. 
Hereafter we use M$_\mathrm{500}$ instead of M$_\mathrm{200}$, to enable a comparison at the same radius with previous studies on $f_\mathrm{b}$ in nearby groups/clusters. 
The catalogue value of M$_\mathrm{200}$ is converted into M$_\mathrm{500}$
assuming an NFW profile with a constant concentration parameter ($c=$5).

\subsection{COSMOS galaxies: multiwavelength photometry and photometric redshifts}\label{redshifts}

The COSMOS area has been imaged in 30 bands including broad- (SUBARU \citealt{taniguchi07a}; CFHT McCracken et al. in preparation), medium-, and narrow-bands (SUBARU; Taniguchi et al. in preparation), ranging from the far-ultraviolet (GALEX \citealt{zamojski07}) to the mid-infrared ({\it Spitzer} \citealt{sanders07}). 
This multiwavelength dataset is collected in a master photometric catalogue. 
Capak et al. (in preparation) discuss in detail source detection and extraction of photometry.
The COSMOS photometric catalogue is complete down to a total $i$-band magnitude
of 26.5 AB mag.
 \citet{ilbert08} and \citet{salvato08} computed  highly reliable photometric redshifts with unprecedented accuracy for a survey this large, owing to the extraordinarily large number of photometric bands.
Redshifts were attributed to individual galaxies
via a standard $\chi^2$ fitting procedure \citep{arnouts02}
encoded in {\it Le Phare}\footnote{www.lam.oamp.fr/arnouts/LE\_PHARE.html},
written by S.~Arnouts and O.~Ilbert.
Best-fit solutions from this photo-$z$ algorithm
were trained on a composite spectroscopic sample of objects brighter than  $i_\mathrm{AB} =25$ \citep[see table 3 in ][]{ilbert08}, mostly made of $\sim$4,000 bright galaxies (i.e., with $i_\mathrm{AB} < 22.5$) observed as part of the $z$COSMOS spectroscopic survey \citep{lilly07}.
Comparison of photometric and spectroscopic redshifts
gives a typical r.m.s. scatter of the photo-$z$'s
equal to $\sigma_{\mathrm{photo}-z} = 0.02 \times (1+z)$
for $i_\mathrm{AB} \le 25$ and $z < 1.25$  \citep{ilbert08}.
In the presence of X-ray emitting objects (AGNs), photometric redshifts were independently estimated
by \citet{salvato08}.

As a by--product of the photo-$z$ determination,
spectroscopic types were attributed to individual galaxies
on the basis of their best-fit broad-band spectral energy distributions (SEDs).
This information is used to estimate the stellar mass of a galaxy,
which is obtained from the conversion of the Ks-band luminosity
\citep{ilbert08} using an evolving galaxy--type dependent stellar mass-to-Ks-band luminosity ratio
$M/L_\mathrm{Ks}$ \citep{arnouts08}. This relation  
has been established using a Salpeter initial mass function \citep{salpeter55}.
Stellar masses of individual galaxies are contained
in the COSMOS photometric catalogue; 
the fractional error on the stellar mass of a galaxy is typically  
	equal to 34$\%$ , and is dominated by the mean scatter on $M/L_{Ks}$ \citep{arnouts08}.  
	
This uncertainty pertains to the aforementioned method of estimating stellar masses. 
Individual galaxy stellar masses may differ by a factor 2--3, depending on the method used to estimate the mass (e.g \citealt{longhetti08}; \citealt{aybuke07}). 
This uncertainty is the product of several factors;
it mostly reflects the range of assumptions in differing models
as for the star--formation history (e.g., single burst
vs. multiple bursts vs. continuum star-formation activity)
and the attenuation of stellar light by dust
(e.g., starburst-like vs. normal star-forming disc-like).
In addition, it results from different implementations of complex physics, such as
the asymptotic--giant--branch phase of stellar evolution
and metal enrichment). This scatter does not reflect the uncertainty of the present method, which is  34$\%$ for individual galaxies as detailed above. This latter value is the uncertainty we attribute to individual galaxy stellar masses in the present study.

\subsection{Nearby clusters}
The COSMOS sample is mostly composed of groups. Therefore we complement it with a sample of 27 nearby X--ray selected clusters with sufficiently deep 2MASS photometry (\citealt{lin03}, LMS03) to estimate accurate stellar masses. The total and stellar masses were derived by LMS03 in a manner consistent with ours. In particular, the total cluster mass is estimated from an M$_\mathrm{500}$--T$_\mathrm{X}$ relation. The stellar masses are estimated from the total K band luminosity of each cluster, assuming an average stellar  mass--to--light ratio which takes into account the varying spiral galaxy fraction as a function of the cluster temperature.\\
LMS03 provide estimates of the total gas fraction obtained from either X-ray data or from a scaling relation; we use instead the most recent scaling relations of \citet{pratt08}, based on hydrostatic mass estimates, in order to reduce systematic effects. We apply this both to our sample and the one of LMS03.

\section{Data Analysis}\label{analysis}

\subsection{Galaxy stellar mass function: completeness and extrapolation}\label{galaxymassfunction}

	The low--mass end of the galaxy stellar mass function
	of the individual COSMOS groups/poor clusters
	is probed to different extents by observations,
	since these systems span a rather large redshift range ($0.1 \le z \le 1$).
	In order to achieve a common footing, the completeness in galaxy absolute magnitude
	(stellar mass) of the sample must be understood.

\begin{figure}[!h]
\begin{center}
\includegraphics[width=\columnwidth]{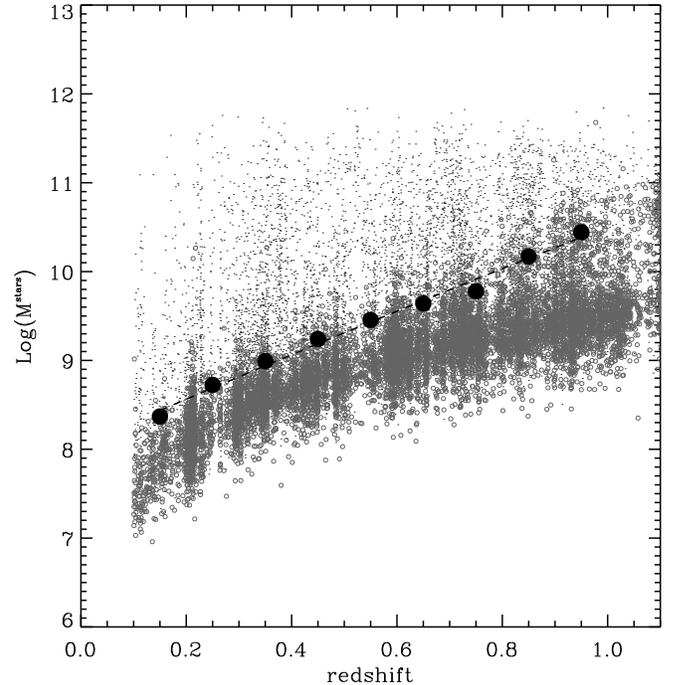}
\caption{The completeness stellar mass for our sample is computed from the fit (black dashed line) to the 95$\%$ percentile of the distribution in $M_\mathrm{lim}$ (see text) for galaxies in the 20$\%$ lower percentile in magnitude (grey circles) as a function of redshift. The black dots represent the stellar masses for all galaxies with $i_{AB}\le 25$. To reduce the plot size, we plot only one point in ten.\label{fig2}}

\end{center}
\end{figure}

	First, we divide the sample into two redshift bins (0.1--0.5 and 0.5--1.0) containing a similar number 	of objects, since the cosmic stellar mass density is observed  to drop by a factor of 2 from $z\sim$0 to 1 in the field (\citealt{wilkins08} and references therein).
	The completeness mass is estimated at $z$=0.5 and $z$=1.0 from a fit  of its behaviour as a function of redshift, obtained using a sampling of 0.1 in redshift as follows (Bolzonella et al. in preparation). Firstly we derive the stellar mass ($M_\mathrm{lim}$) that each object would have if its apparent magnitude was equal to the sample limit magnitude (i.e. $i_{AB}$=25), viz. , 
	\begin{equation}
\log M_\mathrm{lim}=\log{M}+0.4\times (i_{AB}-25.0) , 
\end{equation}
where M is the stellar mass of a galaxy with apparent magnitude $i_\mathrm{AB}$.
Secondly we derive the 95$\%$ percentile of the distribution in $M_\mathrm{lim}$ for galaxies in the lower 20$\%$ percentile in magnitude (i.e. $i_{AB}\ge 23.6$) in each bin of 0.1 in redshift. Finally a fit to the corresponding envelope as a function of redshift  is performed for 0.1$\le z \le$1.0; the ensuing values represent the stellar mass completeness  as a function of redshift for our sample.  Figure \ref{fig2} illustrates the behaviour of the stellar mass completeness as a function of redshift.

For instance, the stellar mass completeness 
 	at z=1 (M$_\mathrm{compl}=10^{10.4}$M$_{\odot}$) is about an order of magnitude lower
	than the so-called ``transition'' stellar mass at $z \le 1$
	\citep[e.g.][]{bundy05, pannella06}.
	This confirms that a rich mixture of morphologies
	and, thus, star-formation histories \citep{sandage86} is present
	among the member galaxies of the COSMOS X-ray selected groups/poor clusters.

	We compute the total stellar mass associated with galaxies of a given system as follows. We first add 
	 the stellar masses of galaxies
	more massive than the completeness mass (at $z$=0.5 or 1)
	for which membership to a given group/poor cluster is determined
	(as described in $\S$\ref{membership}). Taking into account the mass of the individual galaxies, rather than their statistical distribution (as in \citealt{lin03}), becomes increasingly important for groups, where the BCG is a large fraction of the total stellar mass.
	
		The contribution from less massive galaxies is estimated in a statistical manner from the composite stellar mass function (Giodini et al. in preparation), which can be robustly obtained only within two broad redshift bins (0.1$\le z \le$0.5 and 0.5$< z \le$1.0). 
		The stacked stellar mass function for systems falling in each redshift bin is fitted with a single Schechter function \citep{schechter76}; the correction factor for stellar masses lower than the completeness mass, down to $\sim$10$^8$ M$_{\odot}$ (typical mass of a dwarf galaxy), is given by:
		\begin{equation}
 		1-\frac{\int^{10^{13}}_{\mathrm{M}_\mathrm{compl}} f({M})\cdot {M}\,\mathrm d{M}}{\int^{10^{13}}_{10^{8}} f({M})\cdot \,\mathrm d{M}}, 
		\end{equation}
	where M$_\mathrm{compl}$ is the completeness mass for the given redshift range.
	The fractional contribution to the total stellar mass budget of galaxies with $10^{8}~ \mathrm{M_{\odot}} \le$ M$  \le$ M$_\mathrm{compl}$
	corresponds to $\sim$9$\%$ ($\sim$1$\%$) at redshifts 0.5--1.0 (0.1--0.5).		These values are almost negligible, as in \citet{lin03}, which confirms
	that the total stellar mass associated with galaxies
	can be achieved almost directly from the data for our sample of X-ray-selected groups/poor clusters at $0.1 \le z \le 1$.

\subsection{Total stellar mass (in galaxies)}\label{stellarfraction}

	\subsubsection{Statistical membership}\label{membership}

	As a first step, we estimate a projected total stellar mass, which is the sum of the stellar masses of all potential member galaxies down to the completeness mass of either redshift bin to which a group belongs (i.e.  0.1--0.5 or 0.5--1.0).  Candidate members are defined as 
	all the galaxies within a projected distance equal to R$_\mathrm{500}$
	from the X-ray centroid of a group/poor cluster and within $0.02 \times$(1+z) from its redshift (given in the X-ray catalogue).
	Then we perform a foreground/background correction by
 	measuring the total stellar mass of galaxies
	contained in 20 circular areas which have the same radius as R$_\mathrm{500}$ and have photometric redshifts consistent with that of a given system within the errors. These areas do not overlap either with the group or with other groups at the same redshift and are chosen to represent the coeval field environment.
	Field galaxies are selected in redshift and stellar mass
	following the same criteria as for the selection
	of potential member galaxies previously described.
	The mean and the standard deviation
	of the distribution of the total stellar masses computed in the 20 regions
	are taken as the value of the stellar mass associated with the foreground/background
	and its uncertainty, respectively.
	Finally, the foreground/background value
	is subtracted from the initial estimate of the total stellar mass of the system.\\ 
	If the error on the foreground/background value is larger than half
	of the estimated total stellar mass content in galaxies of a given system,
	this system is removed from the sample.
	Obviously a system is excluded also if the foreground/background correction
	exceeds the estimated total stellar mass content in galaxies. The variance on the total stellar mass budget in galaxies for a system
	is given by the sum in quadrature of the background uncertainty
	and the error on the total stellar mass of the galaxies of the system.
	
		Furthermore we checked the influence of masked areas on the reliability of the computed total stellar masses of individual groups. 
	A region of the COSMOS area is masked when the image quality is poor owing to different reasons (e.g. field boundary, saturated stars, satellite tracks and image defects).
	For galaxies with elliptical-like SEDs reliable photo-$z$'s can be determined also in masked areas; therefore early--type galaxies falling in masked areas are considered. On average, the contribution of these objects to the stellar mass budget of a group is not expected to be negligible.  In fact, in 30 out of 37 cases where early--type galaxies falling in masked areas are retrieved, the new stellar mass fraction (computed in $\S$\ref{star} ) is consistent with that of other groups with the same M$_\mathrm{500}$, whatever the redshift.
Conversely, late--type galaxies falling in masked areas are not considered and the impact of this 		choice is tested {\it a posteriori}.
	 For 23 out of 114 groups the number of statistically established member galaxies is less than 6 and the total stellar mass is systematically lower than the mean for groups of similar total masses, irrespective of M$_\mathrm{500}$ \footnote{This tells that 5 members only is insufficient to determine the stellar mass budget of a group. 
	In fact, when the total stellar mass or luminosity of a system
	is computed from a population of discrete sources,
	the scatter in the ensuing value turns out to be non linear
	when the number of discrete sources becomes small (e.g.  of order ten or less),
	as demonstrated by \citet{gilfanov04} in an analogous application.}.
 These 23 objects span the entire total mass range and their exclusion does not affect our results on the stellar mass fraction; at the same time, the scatter in the stellar mass fraction decreases by 30$\%$ \footnote{Nevertheless these objects are potentially an interesting sub-population characterized by an extremely slow build-up of stellar mass. Further optical follow up will help to better assess  their properties.}.	
	 Only the resulting sample of 91 galaxy systems with at least 6 members, spanning two orders of magnitude in X--ray luminosity,  is considered in the following analysis; hereafter it is designated the COSMOS X-ray selected group sample.	
	
\subsubsection{Deprojection}\label{deprojection}

\begin{figure}[!h]
 		\centering
 		\includegraphics[width=\columnwidth]{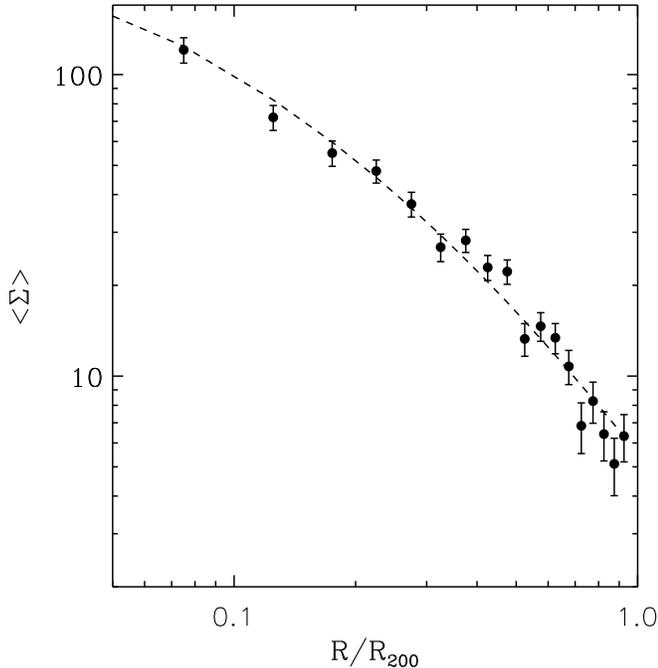}
		\caption{Radial profile of the average number galaxy density for the 91 COSMOS groups/poor clusters. The dashed line shows the best fit NFW profile (c$\sim$4). The unit of the surface density is number per area in unit of $\pi$R$_{200}^2$ and normalized to the total number  of systems.\label{fig3}}
 		
		\end{figure}

	The total stellar mass in galaxies so far estimated
	refers to a cylindrical section of the system
	projected onto the plane perpendicular to the line of sight.
	We therefore need to deproject the total stellar mass
	from two to three dimensions.
	The average galaxy distribution is described by a projected NFW profile
	in two dimensions \citep{bartelmann96, navarro97}:

		\begin{equation}
  		\Sigma(x) = {2\rho_{\rm s}r_{\rm s}\over x^2 - 1}\,f(x)\;,
		\end{equation}
	where
		\begin{equation}
 		f(x) = \cases{
  		1 - {2\over\sqrt{x^2 - 1}}\arctan\sqrt{x-1\over x+1} & $(x>1)$ \cr
  		1 - {2\over\sqrt{1 - x^2}}\arctanh\sqrt{1-x\over 1+x} & $(x<1)$ \cr
  		0 & $(x=1)$ \cr
  		}
		\end{equation}
	and as a generalized NFW profile in three dimensions
		\begin{equation}
 		\rho(x) = {\rho_{\rm s}\over x(1+x)^2}\;.
		\end{equation}
	In both equations the radial coordinate $x$ is the radius in units of a scale radius $r_{\rm s}$, $x		\equiv r/r_{\rm s}$. The scale radius corresponds to the ratio between R$_\mathrm{200}$ and the 			concentration parameter $c$ for the system.
	An average profile is produced using all 91 systems in our final sample, with a central density 			normalized to the number of groups. This high signal-to-noise, average two-dimensional galaxy distribution
	is best-fitted by a two-dimensional NFW profile where $r_\mathrm{s}=0.27\,R_{200}$.
	The average radial profile is shown in Figure \ref{fig3} together with its best fit (with a reduced 		$\chi^2$ value equal to 1.2). We remark that our aim is not to compute the concentration parameter 	of the galaxy distribution for individual systems,  otherwise we should take into account the scatter 		in the evolution of the concentration parameter as a function of redshift.
	Instead we want to compute an average correction for projection of the mass profile of a 		system as calculated in $\S$\ref{membership}.

	Using the best-fit values, we compute correction factors by integrating the average profile out to R$_\mathrm{500}$:
		\begin{equation}
		 dpf=\frac{\int^{R_\mathrm{500}}_0 \rho(r)\cdot4\pi r^{2}dr}{\int^{R_\mathrm{500}}_0 \Sigma(r)\cdot2\pi r\,dr}.
		\end{equation}
	 The deprojected total stellar mass of a system is then given by
		\begin{equation}
		\mathrm{M}^\mathrm{stars}_{500}=dpf\times M^\mathrm{stars}_\mathrm{proj,500},
 		\end{equation}
	where $dpf$= 0.86 is the correction factor.
		
\section{Results}

	\subsection{Stellar mass budget (galaxy component)}\label{star}

\begin{figure}[!h]
	 \centering
	 \includegraphics[width=\columnwidth]{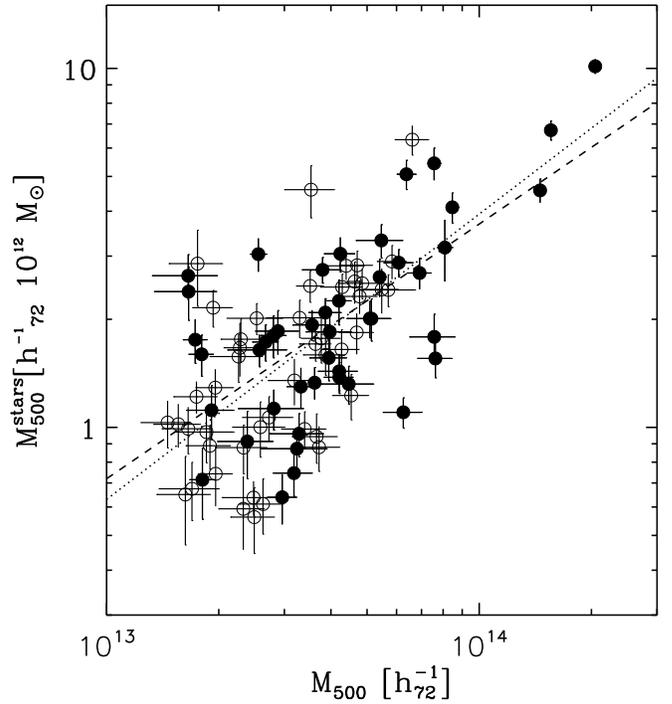}
	 	 \caption{Total stellar mass in galaxies vs. total mass
	for the 91 COSMOS X-ray selected groups/poor clusters. Filled (empty) grey circles identify objects with flag=1 		(2). The dashed (dotted) line represents the best fit relation derived for flag=1 (all) groups (see equation \ref{eq:pivot}) derived taking into account uncertainties in both quantities and the intrinsic scatter of the relation.  }
 	\label{fig5a}
	\end{figure}

	Figure \ref{fig5a} shows the behaviour
	of the total (deprojected) stellar mass in galaxies within R$_\mathrm{500}$, 
	M$^\mathrm{stars}_\mathrm{500}$, as a function of the total mass M$_\mathrm{500}$
	for the 91 COSMOS X-ray selected groups.
	The distribution in Figure \ref{fig5a} exhibits a rather well defined trend,
	although a large scatter is present, especially at low masses,
	where values can range by a factor of 10 at a fixed total mass.
	Part of this large scatter may have a physical origin:
	different merging histories produce different total mass-to-light ratios
	for fixed total assembled mass \citep[cf.][]{sales07}.
 
	We fit the relation between total stellar mass in galaxies and total mass
	for all 91 systems and for the 45 flag=1 groups only.
	Since the distribution in Figure \ref{fig5a} exhibits an intrinsic scatter
	larger than the errors on the individual points,
	the fit is performed using the weighted least square with intrinsic scatter (WLSS) method discussed in \citet{pratt06}.
	This algorithm takes into account uncertainties on both stellar mass and total mass and the presence of intrinsic scatter in the data.
 	There is a robust correlation
	between M$^\mathrm{stars}_{500}$ and M$_\mathrm{500}$ in the COSMOS X-ray selected groups: 

	\begin{equation}\label{eq:pivot}
 M^\mathrm{stars}_\mathrm{500}= (0.30\pm0.02)\times \left( \frac{M_\mathrm{500}}{5\times10^{13}\,h_{72}^{-1}} \right)^{\alpha} \,,
	\end{equation}
	
	where $\alpha$=0.81$\pm$0.11 for the entire sample and $\alpha$=0.72$\pm$0.13 for the flag=1 	subsample, and the (logarithmic) intrinsic scatter  is equal to 35$\%$ in both cases\footnote{This result is robust
	against the presence of a pair of groups which are detected
	at the same redshift, but with a separation of the order of R$_\mathrm{500}$. The two objects of this pair lie
	above the best-fit relation reproduced in Figure \ref{fig5a},
	perhaps as an effect of a bias in their estimated total stellar masses
	in galaxies. However, new fits performed after excluding these two groups
	give the same results as the previous ones.}.

	\begin{figure*}
	\centering
	 \includegraphics[width=1.05\columnwidth]{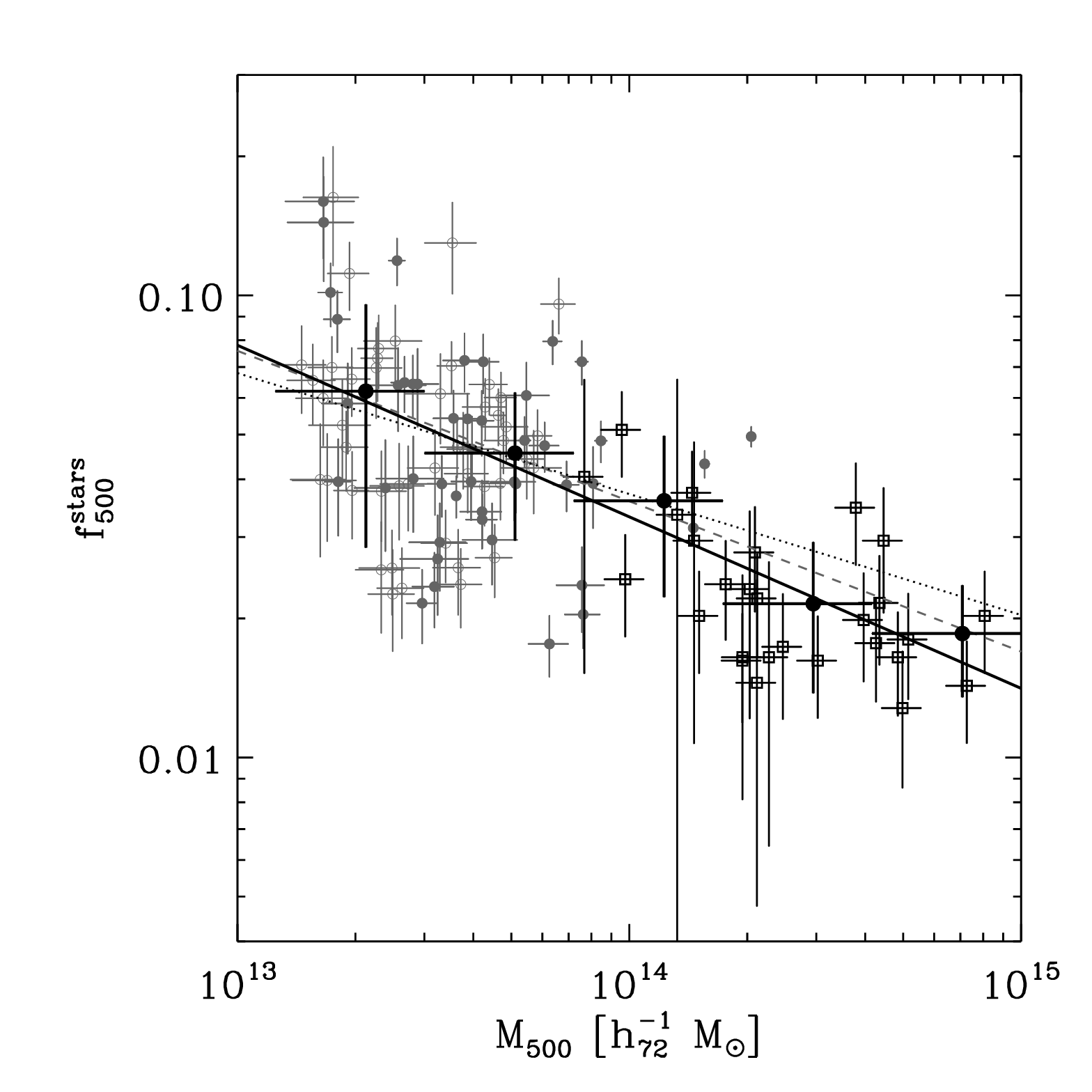}
	  \includegraphics[width=0.9\columnwidth]{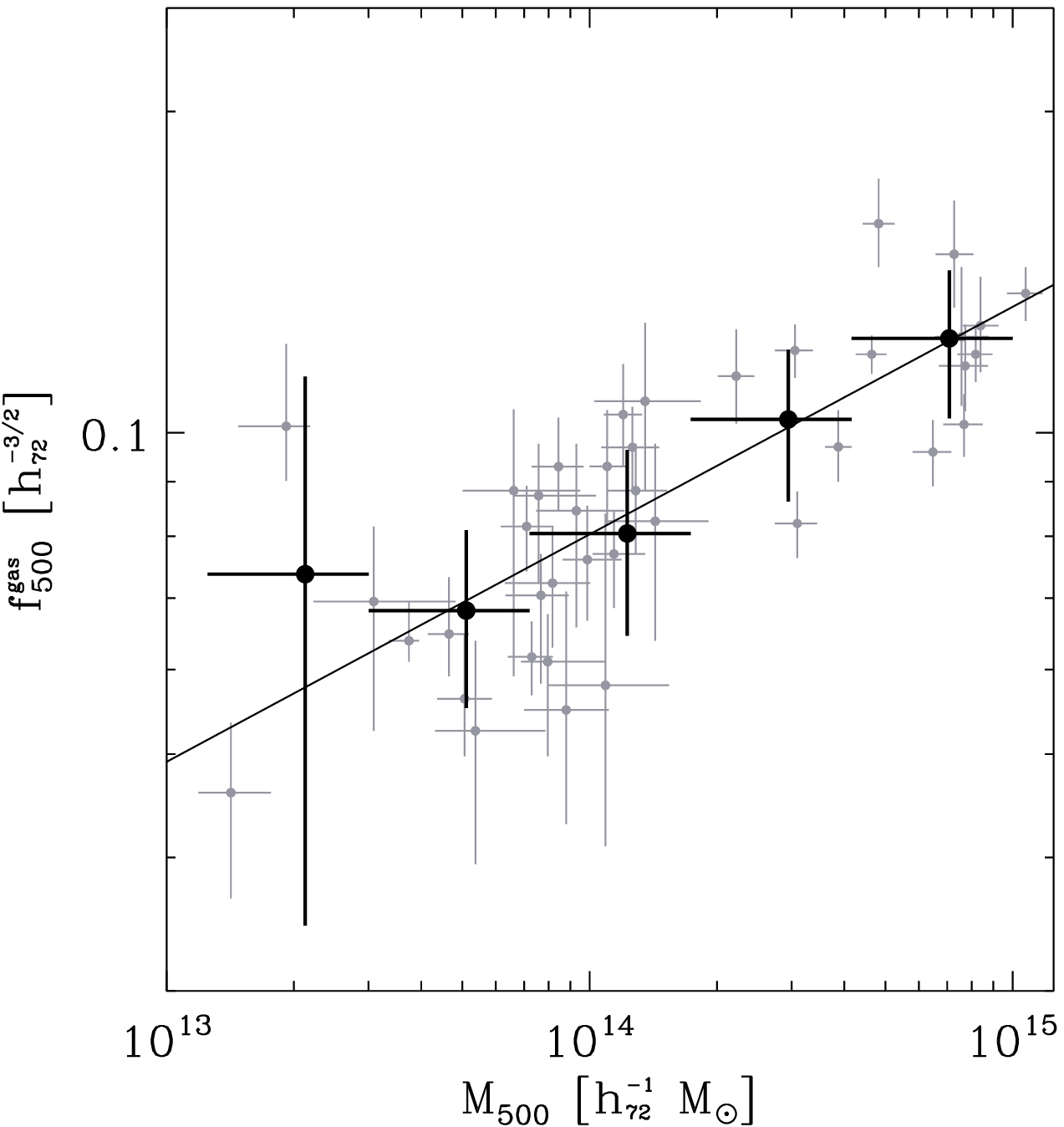}
	 \caption{{\it Left panel}: stellar-to-total mass ratio vs. total mass for the combined sample of  91 COSMOS X-ray selected groups (same symbols as in Figure \ref{fig5a}) plus 27 nearby clusters of LMS03 (empty squares). The dashed line represents the best--fit relation derived for flag=1 groups of the COSMOS sample and the dotted line represents the fit to all COSMOS groups. The solid line shows the best fit relation for all COSMOS groups plus local clusters. All fits are derived taking into account uncertainties in both quantities and the intrinsic scatter in the relation. The ensuing fit parameters are given in Table \ref{tab1}. The large points with error bars show the biweight mean and standard deviation of these data binned in 5 logarithmic bins in total mass. {\it Right panel}: gas fraction as a function of the system mass from a combined sample of 41 clusters and groups (\citealt{vikhlinin06}, V06; \citealt{arnaud07},  APP07; \citealt{sun08}, S08). The solid line is the best fit relation $f^{\rm gas}_{500}\propto M_\mathrm{500}^{0.2}$. The large points with error bars show the mean and standard deviation of these data binned in 5 bins of total mass.}
	  \label{fig5b}
	\end{figure*}

	Fitting the stellar-to-total mass ratio vs. total mass of the system for the full sample of COSMOS X-ray selected groups  only  we find 
		\begin{equation}\label{smass1}
 	f_\mathrm{500}^\mathrm{stars}=5.0^{+0.1}_{-0.1}\times10^{-2}\,\left(\frac{M_\mathrm{500}}	{5\times10^{13}\,M_{\odot}}\right)^{-0.26\pm0.09} .
	\end{equation}
	A fit to the Flag=1 sample gives equivalent results.
	Remarkably the relation between the mass fraction of stars in galaxies and the total mass of the system for the COSMOS X--ray selected groups is consistent within the errors with the one found in nearby clusters by LMS03 and \citet{lagana08}.
We now extend the range of total masses using the results from local clusters selected by LMS03, converting their measurements to our cosmology. Since these authors do not give the uncertainties associated with their total mass estimates, we assign a fixed fractional total mass uncertainty equivalent to the mean of that for the COSMOS groups ($\sim$30$\%$). The best fit of the combined sample is
		\begin{equation}\label{smass2}
 	f_\mathrm{500}^\mathrm{stars}=5.0^{+0.1}_{-0.1}\times10^{-2}\,\left(\frac{M_\mathrm{500}}	{5\times10^{13}\,M_{\odot}}\right)^{-0.37\pm0.04} ,
	\end{equation}
	with a typical logarithmic intrinsic scatter of $\sim$50$\%$. The data and best fit relations are shown in  Figure \ref{fig5b}. 
	\begin{deluxetable}{lcc}[!h]
	\tabletypesize{\small}
	\tablewidth{0pt}
	 \tablecaption{The best fit parameters  for the relation between stellar mass fraction  and total mass (Eq. \ref{smass1} and Eq. \ref{smass2}) for three samples considered. Data were fitted with a power law  $f^{\rm stars}_{500}=N(M_\mathrm{500}/5\times10^{13}\,M_{\odot})^{\alpha}$.\label{tab1}}
	
	 \tablehead{
	 \colhead{Sample} &\colhead{Log(N)\tablenotemark{a}} & \colhead{slope}
	 }
	  \startdata
	COSMOS flag=1 & -1.35$\pm$0.01 & -0.33$\pm$0.12 \\
	COSMOS flag=1+2 &-1.35$\pm$0.01& -0.26$\pm$0.09 \\
	COSMOS+LM03 & -1.37$\pm$0.01 & -0.37$\pm$0.04 \\ 
	\enddata
	  \end{deluxetable}
	   To better elucidate trends with total mass, we divided the data set
	into five logarithmic bins of equal size in total mass, and computed the mean and
	standard deviation of the values of the mass fraction of stars in galaxies in each bin using the biweight estimators of \citet{beers90}; they are relatively large, which gives a measure of the heterogeneity of the population. The large
	points with error bars show the trend of these binned data with total
	mass: there is good agreement with the best fitting regression line to the unbinned points, as
	expected.
	
 \begin{deluxetable}{cccc}[!h]
\tabletypesize{\small}
\tablewidth{0pt}
 \tablecaption{Measured values for $f_\mathrm{500}^\mathrm{gas}$ and $f_\mathrm{500}^\mathrm{stars}$ as in Figure \ref{fig6}. Uncertainties correspond to the standard deviation of the mean (see text for details).\label{tab2}}

 \tablehead{
 \colhead{M$_\mathrm{500}$/ [h$_{72}^{-1}$M$_{\odot}$]} & \colhead{f$_\mathrm{500}^\mathrm{stars}$} & \colhead{f$_\mathrm{500}^\mathrm{gas}$} & \colhead{f$_\mathrm{500}^\mathrm{stars+gas}$}}
 \startdata
2.1e+13    & 0.062$\pm$0.005	 & 0.074$\pm$0.028    &        0.136$\pm$0.028 \\
5.1e+13	& 0.045$\pm$0.002	&  0.068$\pm$ 0.005   &       0.113$\pm$0.005\\
1.2e+14	& 0.036$\pm$0.004	& 0.080 $\pm$ 0.003   &       0.116$\pm$0.005\\
3.0e+14	& 0.021$\pm$0.002	& 0.103  $\pm$0.008   &	    0.124$\pm$0.009\\
7.1e+14   & 0.019$\pm$0.002	& 0.123  $\pm$0.007  &	   0.141$\pm$0.007\\
\enddata
 \end{deluxetable}	 
		\subsection{Evolutionary considerations}
	Finally we inspect the presence of evolution of the relation between $f_\mathrm{500}^\mathrm{stars}$ and M$_\mathrm{500}$ by considering only systems at z$\le$0.5 (we cannot fit the relation for the high redshift systems since they do not cover a sufficient range in total mass). The ensuing fit is fully consistent with that obtained for the entire sample within the uncertainties. 
	
 We can put a constraint on the possible evolution of the relation by evaluating the change in the mean of $f_{star}$ for massive systems (M$_{500}>5\times10^{14}$M$_{\odot}$) in two redshift bins (z$\leq$0.5 and z$>$0.5). The average $f_{star}$ changes from 0.031$\pm$0.013 at  z$\leq$0.5 to 0.039$\pm$0.019 at z$>$0.5, a less than one sigma difference in mean values. Even taking the maximum distances between the two values given the uncertainties, the stellar mass fraction does not change by more than 35$\%$.

A second way to investigate a possible evolution of the stellar mass fraction in galaxies is to plot the ratio of the stellar fraction to the mean relation as a function of redshift (r$_{f}(z)=f_{star}(z)/\langle f_{star}\rangle$). Using the same five bins in total mass as above, no trend in r$_{f}$(z) is evident. However a fit of r$_{f}$(z) gives a robust upper limit on the evolution over the maximum redshift range (0--1) of 40$\%$. Taking the median redshift of each redshift bin (0.22, 0.72), the upper limit on the evolution of the stellar fraction is less than 20$\%$.  This number is consistent with the upper limit on the evolution of the relation between total star fraction and M$_{500}$ given by \citet{balogh08}. 
Therefore we conclude that our data do not support the existence of a significant evolution in the zero--point and slope of the $f_\mathrm{500}^\mathrm{stars}$--M$_\mathrm{500}$ relation between redshifts 0 and 1. 

\subsection{The total baryon mass fraction}
\subsubsection{The gas mass fraction}
	In order to determine the total baryon mass fraction in individual systems,
	we need to estimate the amount of baryons in the form of hot gas which make the intra--cluster medium (ICM). Unfortunately, this cannot be achieved from most of the existing X-ray observations
	of the total sample because their signal--to--noise is insufficient for the purpose.
	Therefore, we have to resort to an estimate of the mean trend of the gas mass fraction as a function of M$_{500}$ 
	established from an independent sample of well observed groups and clusters
	at $z \le 0.2$, selected from the samples of \citealt{vikhlinin06} (V06), \citealt{arnaud07} (APP07) and \citealt{sun08} (S08).
These authors computed gas mass fractions at R$_\mathrm{500}$ from hydrostatic mass estimates for 10  (V06), 10 (APP07) and 21 (S08, including the best quality tiers 1 and 2 systems) clusters and groups, 		respectively. The combined sample contains 41 systems and spans the total mass range 1.5$	\times$10$^{13}$-1.1$\times$10$^{15}$ M$_{\odot}$. After conversion to a common cosmology, a 		fit of the combined data set using the WLSS regression yields:
		\begin{equation}
f^\mathrm{gas}_{\rm500} (h/0.7)^{3/2}= (9.3^{+0.2}_{-0.2})\times 10^{-2} \,\left(\frac{M_\mathrm{500}}	{2\times10^{14}\,M_{\odot}}\right)^{0.21\pm0.03} .
		\end{equation}
	\noindent with a scatter of 17 per cent about the best fitting
	regression line. The data and resulting fit are shown in
	Figure.~\ref{fig5b}.
	As discussed in the introduction we assume that this relation is not evolving, in the absence of observations to the contrary.
	To better elucidate trends with total mass, we divided the data set
	into the same logarithmic bins in total mass as for the stellar mass fraction, and computed the mean and standard deviation of the distribution of the gas mass fraction values in each bin. The large
	points with error bars show the trend of these binned data with total
	mass. The observed relation suggests that lower mass systems have proportionally less gas than high mass systems. Further discussion is available in \citet{pratt08}.

\subsubsection{The baryon mass fraction (in galaxies and ICM)}\label{baryonic}
We now combine the results on the stellar and gas mass fractions derived in the previous two sections to investigate the behaviour of the baryonic mass fraction as a function of total mass. At this stage no contribution is considered from the ICL as defined in $\S$\ref{ICL}.
In each logarithmic mass bin we sum the mean contribution from stellar and ICM mass components. As we wish to determine the behaviour of the average systems in a given mass bin, for each component the uncertainty is calculated from the standard deviation of the mean (the standard deviation divided by $\sqrt{N-1}$, where $N$ is the number of data points in the bin). The uncertainty on the total baryon mass content is then estimated from the quadratic sum of the individual uncertainties for the stellar and ICM contributions.
	Figure \ref{fig6} (lower panel) reproduces the average behaviour of the sum of the two baryonic components estimated in the previous sections (i.e. ICM gas and stars associated with galaxies) as a function of total mass for galaxy systems with 2$\times$10$^{13}\le$M$_{500}\le$8.1$\times$10$^{14}$ M$_{\odot}$. 
	The ensuing baryon mass fraction is an increasing function of the system mass:
	\begin{equation}
		 f_{500}^\mathrm{stars+gas}=(0.123\pm0.003)\times\left(\frac{M_\mathrm{500}}{2\times10^{14}\,M_{\odot}}\right)^{0.09\pm0.03} ,
	\end{equation}
	This expression is obtained after excluding the lowest mass point which is affected by an extremely large uncertainty since the corresponding gas fraction is estimated from only two groups.
	
	\subsection{Comparison with WMAP}\label{comparison}
	\subsubsection{Raw values}
	As Figure \ref{fig6} shows, there is a gap between the values of $f_\mathrm{500}^\mathrm{stars+gas}$ estimated from WMAP5 and those obtained here; this discrepancy, before any correction, is significant at more than 5$\sigma$ for systems less massive than $\sim$10$^{14}$M$_{\odot}$ (see Table \ref{tab3}), where the uncertainties are calculated as described in $\S$\ref{baryonic}. 
	\subsubsection{Values corrected for gas depletion}
	 We now correct the value of the baryon fraction for gas depletion. As discussed in \citet{frenk99}, simulations without feedback suggest that the ICM has a slightly more inflated distribution than the dark matter (see also observations by \citealt{pratt02}), resulting in a decrease in the gas fraction of 10$\%$ at R$_{500}$. In the absence of indications to the contrary we do not assume a mass dependence for the gas depletion.
		 For average massive clusters ($\langle$M$_{500} \rangle=7\times 10^{14} $M$_{\odot}$) the value of gas depletion--corrected $f_\mathrm{500}^\mathrm{stars+gas+depl}$ is consistent within 1.4$\sigma$ with the WMAP5 estimate. However the gas depletion corrected value in the group regime ($\langle$M$_{500} \rangle=5\times 10^{13} M_{\odot}$) is still 4.5$\sigma$ discrepant from that of WMAP5\footnote{We note that this discrepancy represents a lower limit if a further 10$\%$ reduction of the gas mass is applied due to the clumpiness of the ICM as in \citet{lin03}. However this correction is not applied in most of the studies of gas component in clusters.}.
	 \subsubsection{Values corrected for gas depletion and ICL}\label{ICL}
\begin{figure}
 \centering
 \includegraphics[width=\columnwidth]{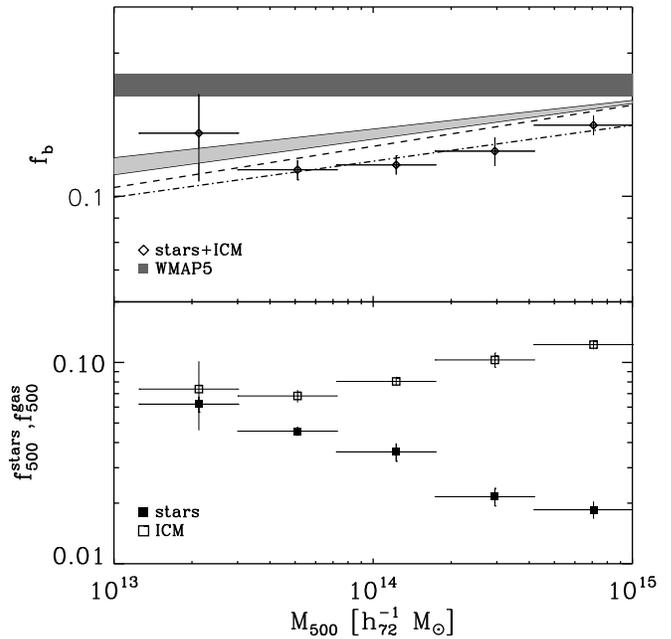}
 \caption{Lower panel: average stellar to dark mass ratio (filled points) for the COSMOS+LM03 sample and average gas fraction (empty points). Uncertainties are computed from the standard deviation of the mean in all cases. Upper panel: total baryonic fraction obtained summing the points in the lower panel compared with the universal value by WMAP5 (dark grey stripe). The dashed-dotted line represents the fit to the measured points. The dashed line represents the fit to the points where the gas fraction has been corrected for a 10$\%$ gas depletion. The light grey stripe is the fit to the relation taking in account both gas depletion and a constant (11--22$\%$) ICL contribution to the stellar mass. }
 \label{fig6}
\end{figure}

	 The existence of a diffuse stellar component in galaxy groups/clusters is now a well established observational result, but the way the ICL is defined and measured is not unique (see \citealt{zibetti08} for a recent review). 
	The quality of our observations is insufficient to measure the contribution of diffuse, very low surface brightness light ($>$25.8 K--mag arcsec$^{-2}$) within r$_{500}$ directly for individual systems in the sample. 
	To quantify the amount of stellar mass which is associated with diffuse light that escapes detection during the standard photometry extraction with SExtractor \citep{capak07}, we are guided by previous observational results. In particular we consider \citet{zibetti05}, \citet{krick07} and \citet{gonzalez05}.
	\citet{zibetti05} used stacking analysis of 683 systems at z=0.2--0.3 ranging in total mass from a few times 10$^{13}$ to 5$\times$10$^{14}$M$_{\odot}$ (the average total mass is 7$\times$10$^{13}$M$_{\odot}$), selected from a ~1500 deg$^{2}$ of SDSS-DR1, reaching the unprecedented surface brightness limit of $\sim$32 mag arcsec$^{-2}$ (R--band in the z=0.25 observed frame). They show that on average the  ICL contributes $\sim$11$\%$ of the stellar light within 500 kpc.
	In a complementary study \citet{krick07} used a sample of massive clusters with a range of morphology, redshift and densities to find that the ICL contributes with 6$\%$--22$\%$ to the total cluster light in $r$--band within one quarter of the virial radius, finding no appreciable correlation with cluster mass. 
	Given these results, we assume that the contribution of the ICL to the total mass of a system is equal to its observed contribution to the total light and ranges between 11 and 22$\%$. This range is consistent with the theoretical results by \citet{murante07} and \citet{purcell08}, in their attempt of modelling the ICL by numerical simulations.
	Furthermore given the complete lack of observational constraints, we assume that the ICL mass fraction is not evolving with redshift for 0$<$z$<$1; this is supported by the simulation of \citet{dubinski03} as shown in \citet{feldmeier04}.
		 We discuss the impact of our choice on the results in $\S$\ref{systematics}.
	  	  The final gas depletion corrected values including the ICL contribution of f$^\mathrm{stars+gas+depl+ICL}_\mathrm{500}$ are lower than the WMAP5 estimate across the entire explored mass range;  f$_\mathrm{500}^\mathrm{stars+gas+depl+ICL}$ is in agreement with the WMAP5 result within 1$\sigma$ in the massive cluster regime, but still discrepant at a significance level of at least 3.3$\sigma$ for groups (see Figure \ref{fig6}). 
		 	
 \begin{deluxetable}{cccc}
\tabletypesize{\small}
\tablewidth{0pt}
 \tablecaption{Discrepancy of $f_\mathrm{b}$ from the WMAP5 value in sigma units\label{tab3}}
\tablehead{
 \colhead{M$_\mathrm{500}$/ [h$_{72}^{-1}$M$_{\odot}$]} & \colhead{$\Delta_{f_{b}}$/ [$\sigma_{f_{b}}$]} &\colhead{$\Delta_{f_{b}}$/ [$\sigma_{f_{b}}]$\tablenotemark{a}} & \colhead{$\Delta_{f_{b}}$/ [$\sigma_{f_{b}}$]\tablenotemark{b}} }
 \startdata
2.1e+13    & $>$1.2	& $>$0.8  &$>$0.3 \\
5.1e+13	&   5.3 & 4.5  &3.3 \\
1.2e+14	& 5.1 &4.2  & 3.2\\
3.0e+14	& 3.7 &2.6 & 2.1	\\
7.1e+14   & 2.6	  & 1.4& 1.0\\
 \enddata
 \tablenotetext{a}{After correction for gas depletion.}
 \tablenotetext{b}{After correction for gas depletion and ICL.}
 \end{deluxetable}

\subsection{Impact of systematic effects}\label{systematics}
 The basic observational result of the present study is that the baryon mass fraction, corrected for gas depletion and ICL contribution, is consistent with WMAP5 estimate within 1$\sigma$ for clusters with $\langle$M$\rangle$=7$\times$10$^{14}$M$_{\odot}$ but is significantly (3.3$\sigma$) lower for groups with $\langle$M$\rangle$=5$\times$10$^{13}$M$_{\odot}$.
 At the cluster scale our result on the baryon fraction is consistent with that of \citet{lin03}, indicating that different approaches do not show systematic differences in the determination of the gas fraction scaling with the cluster mass.
  Furthermore we note that the scaling relation determined by \citet{pratt08} is based on three different samples of groups and clusters: this should reduce the potential bias produced by sample selection. In \citet{pratt08} the best fit relation to the combined data from hydrostatic estimates reproduces the REXCESS sample distribution where the gas masses have been estimated using the M--T relation of \citet{arnaud05}. This suggests that potential systematic effects on our estimates of the gas mass fractions at low redshifts are negligible.
  
  In the absence of direct estimates of the gas fraction at z$>$0.2, we have to rely upon the results of existing simulations, which predict the gas fraction within r$_{500}$ to increase on average by $\sim$5$\%$ (adiabatic simulations) or 10--20$\%$ (simulations with cooling and star formation) between z=0 and z=1 for groups and poor clusters \citep{kravtsov05}.  Applying a correction to this effect at the median redshift of the COSMOS group sample (by 5--10$\%$ at z=0.5), the discrepancy in the baryon mass fraction between groups of $\langle$M$\rangle$=5$\times$10$^{13}$M$_{\odot}$ and WMAP5 is reduced to 3.0--2.6 $\sigma$. Therefore we conclude that systematic underestimates of the gas fraction alleviate but do not solve  the discrepancy at the group scale. 
  
Since inside groups the stellar mass fraction is comparable to the gas mass fraction, we analyze the impact of the ICL fraction and the adopted stellar mass--to--light ratio (M/L) of the galaxy population.
We have adopted a mass independent correction to the total stellar mass fraction for ICL, equal to 11--22$\%$. If a  strong anti--correlation between the ICL mass fraction and the total mass of the system exists, and the true ICL mass fraction is equal to $\sim$50$\%$ at the group scale, an agreement between our total baryonic mass fraction and the WMAP5 estimate is reached. Such a figure has been claimed by \citet{gonzalez07} for a sample of 23 BCG--dominated clusters and groups. However the ICL--to--BCG light ratio (ICL/BCG) is strongly dependent on the decomposition of the total surface brightness profile of the two components and the photometric depth (\citealt{gonzalez05}; \citealt{zibetti08}). We note that \citet{gonzalez05} give ICL/BCG$>$5 by applying a simultaneous decomposition of the surface brightness distribution of BCG+ICL in two De Vaucouleurs components: the outer one is considered as the genuine ICL and the inner one as the BCG. Conversely, \citet{zibetti05} obtain ICL/BCG$<$0.5 by fitting only the inner profile with a De Vaucouleurs model (which represents the BCG) and considering all the residual light as ICL. Nevertheless, \citet{zibetti08} applied a two--De Vaucouleurs decomposition to the \citet{zibetti05} data obtaining ICL/BCG$\sim$2, and concluded that the ICL+BCG--to--total light ratio is a much more robust measurement, which is likely equal to 0.3 (in light) for systems of average mass 7$\times$10$^{13}$M$_{\odot}$. 
The high value of ICL/BCG found by \citet{gonzalez07} may be the result of a sample bias, as suggested by the same authors. 
On the other hand the lack of trends reported by Zibetti et al. (2005) could be intrinsically biased by the adoption of a fixed metric aperture of 500 kpc, which 
correspond to smaller fraction of R200 for more massive clusters. Given the steeper proÞle 
of the ICL with respect to galaxies, the ICL fraction of more massive clusters could 
be overestimated and a correction for this effect could reconcile these results with the 
negative trend found by Gonzalez et al. (2007), but not with the extreme values of ICL+BCG--to--total light ratio. Generally, it is evident that better determinations of the trends of the ICL with cluster mass and richness are needed.

The COSMOS groups sample contains a whole range of systems, which exhibit a BCG--to--galaxy stellar mass ratio from 0.2 to 0.9. For these groups the estimated ICL+BCG--to--total light ratio for the average group is 0.36, broadly consistent with the generally accepted average value of 0.3 (\citealt{gonzalez07}; \citealt{zibetti08}). This suggests that we are not missing an important contribution of the stellar mass in our analysis, in spite of our definition of ICL. 

Another systematic effect may be introduced by the computation of the stellar mass--to--light ratio for the ensamble of the member galaxies and the ICL. In our case we use M/L values that correspond to the individual star formation histories of individual member galaxies \citep{arnouts08} and we do not make assumption on the M/L of the ICL. Hence  the major source of systematics on the stellar mass--to--light ratio of our galaxies is given by the adopted initial mass function (IMF). For instance, a change from a standard Salpeter to a Chabrier IMF reduces the M/L by 30$\%$ \citep{longhetti08}. This translates into a decrease by 30$\%$ of the stellar mass associated with galaxies which makes the bulk of the total stellar mass in our systems.
There is no compelling reason to abandon the Salpeter IMF \citep{renzini05}, but it is a possibility explored in the literature. \citet{lin03} obtained the stellar mass--to--light ratio for the ensamble of group/cluster member galaxies by folding in a morphological type dependent M/L with the temperature dependence of the spiral fraction; \citet{gonzalez07} assumed that the ICL and all member galaxies share the same stellar M/L, as the one that characterise an early type galaxy. The latter case assumes that the intergalactic stars are homogeneous with the BCG stellar population. However, it has been suggested that the ICL may (also) origin from the stripping of non--BCG galaxies inside the group/cluster (\citealt{purcell08}; \citealt{pierini08}), which are on average bluer than the BCG, especially in groups (\citealt{zabludoff98}; \citealt{weimann06}; \citealt{poggianti06}). 
For example, if the ICL mass-to-light ratio used in \citet{gonzalez07} is overestimated by a factor 2, it translates in tghe systematic overestimation of the baryon mass fraction by 10$\%$. 

This systematic effect has the same amplitude, but opposite sign, of the potential offset applied to the gas fraction--mass relation according to \citet{gonzalez07}. Therefore we conclude that a 3$\sigma$ discrepancy between the baryon mass fraction of groups and the WMAP value holds against major systematic effects on the stellar populations either diffuse or associated with galaxies.

An overestimate of the total M/L is not enough to explain the values of the stellar mass fraction for the lowest mass systems in \citet{gonzalez07} which largely exceed the constraint on the total baryon fraction set by WMAP5 (as also noted in \citealt{balogh08}). A way out is a systematic and large underestimate of the total masses of these systems, as also suggested by \citet{balogh08}\footnote{These objects certainly impact the strongly inverse total mass dependence of the total stellar mass fraction found by \citet{gonzalez07}.}. 

We conclude that a robust estimate of the total mass is crucial for systems with the lowest mass (in our sample $\langle$M$_{500}\rangle\sim$2$\times$10$^{13}$M$_{\odot}$). Our estimates are based on the L$_\mathrm{X}$--M$_\mathrm{200}$ relation established 
via the weak lensing analysis in Leauthaud et al. (in preparation), and exhibit a typical uncertainty of 30$\%$. The use of different total  mass estimators could offer a test of the presence of systematics, but unfortunately this is still hard to achieve for statistical large samples of groups at different redshifts.

\section{Discussion}\label{discussion}

We have investigated if the discrepancy between estimates of the total
baryon mass fraction obtained from observations of the CMB
and of galaxy groups persists when a large, unbiased sample
of well-characterized groups is considered.
The COSMOS 2 deg$^2$ survey meets this requirement, yielding 91 candidate
X-ray groups/poor clusters at redshift $0.1 \le z \le 1$.
In order to extend the span in total mass to two orders of magnitude
(2$\times$10$^{13}<$M$_{500}<$1.2$\times$10$^{15}$M$_{\odot}$),
we consider 27 nearby clusters investigated by \citet{lin03}.
Comparable robust measurements of total mass and total stellar mass
(in galaxies) exist for individual objects of both subsamples,
as shown in the previous sections.
In addition, the same scaling relation is used to estimate the gas mass
fraction in both subsamples.
This enables us to build a joint sample of 118 X-ray selected groups
and clusters at $z \le 1$ for which the importance of systematics
is reduced (see $\S$\ref{sample}).
For this sample, the behaviour of the total stellar mass fraction
as a function of the total mass
can be investigated for a large range in total mass and, for the first time,
in redshift (at least for groups).
The results of our analysis and their impact on the widely accepted
paradigm of the hierarchical growth of structure in the universe
are discussed hereafter.

\subsection{The stellar mass fraction}\label{result1}
We have shown (Figure \ref{fig5b}) that the stellar-to-total mass ratio in COSMOS groups and in 27 local clusters  is anticorrelated with the total mass of the system. This relation is given by $f_\mathrm{500}^\mathrm{stars} \propto M_\mathrm{500}^{-0.37 \pm 0.04}$, which holds also after introducing the mass independent correction for the ICL (see $\S$\ref{comparison}). 
The global trend between $f_\mathrm{500}^\mathrm{stars}$ and  M$_\mathrm{500}$ is consistent with that observed in clusters at z$<$0.3 both by LMS03 and \citet{lagana08} using much smaller samples. We extend their results to the low mass regime by one decade and to higher redshift. 
 
The difference in the number of stars formed per unit of halo mass between groups and clusters has been interpreted in terms of a varying efficiency of the star formation with the total mass of the system \citep[e.g.][]{lin03}. A variation in the star--formation efficiency for systems with virial temperatures $\ge10^7~$K is a result of simulations by \citet{springel03}; it is interpreted in terms of cooling flows being less efficient in shutting off star formation in groups.
An alternative possibility is that clusters are formed not only by merging of groups and smaller clusters but also that they accrete a large fraction of their  galaxies (with a low stellar mass fraction, of the order of 0.01) from the field (\citealt{white91}; \citealt{marinoni02}). 
However after a mass independent correction for the ICL contribution (introduced in $\S$\ref{comparison}), the relation $f_\mathrm{500}^\mathrm{stars} \propto M_\mathrm{500}^{-0.37 \pm 0.04}$ is in agreement with the constraint on the slope set by the hierarchical model of structure formation under the assumption that at least half of the stars in groups were formed by $z=1$ \citep{balogh08}\footnote{We note that a steeper relation is obtained when the strongly inverse mass dependent ICL fraction of \citet{gonzalez07} is used (see \citet{balogh08} for the discussion).}.
This is supported by the apparent absence of evolution for this relation in our sample within the redshift range 0.1--1.
This shows how observational studies such as the present one can improve the constraints on models and foster our understanding of the underlying physical processes.

\subsection{The total baryon mass fraction}
Combining the computed stellar mass fraction with the estimated gas mass fraction derived from the mean local relation in \citet{pratt08}, we find that the gas plus stellar (galaxies) baryon mass fraction increases by $\sim$25$\%$ (from $\sim$0.11 to $\sim$0.14) when the total mass increases by a factor of one hundred. 
After a constant 10$\%$ correction for gas depletion and a further correction for a constant 11--22$\%$ ICL contribution, the value of $f_\mathrm{500}^\mathrm{stars+gas+depl+ICL}$ for an average cluster is consistent within 1$\sigma$ with the cosmic value measured by WMAP, while the $f_\mathrm{500}^\mathrm{stars+gas+depl+ICL}$ found for an average group differs from it at more than 3$\sigma$. Given the heterogeneity of the sample (see e.g. Figure \ref{fig5b}), for some objects the gap between $f_\mathrm{500}^\mathrm{stars+gas+depl+ICL}$ and the WMAP5 value could be negligible or, conversely, statistically more significant for objects in the same bin of total mass, but at the two extremes of the distribution in $f_\mathrm{stars}^\mathrm{500}$. Unfortunately we do not have a measure of the gas mass fraction for individual objects, therefore we focus on the behaviour of the average object.
We did likewise for the ICL by assuming a fixed fractional contribution of 11--22$\%$ across the entire mass range. Possible systematic effects introduced by our definition and estimate of the ICL contribution are discussed in $\S$\ref{systematics}. Here we stress that they do not lead to an anomalously low BCG+ICL contribution to the total mass of the system. Thus the discrepancy at the groups regime in not erased by uncertainties on the stellar mass fraction.
In the absence of evidence for a systematic and relevant underestimation of the gas mass fraction in our systems (see $\S$\ref{systematics}), we interpret the discrepancy as a lack of gas, by 33$\%$, at the group regime.
This may be produced by feedback (stellar and/or AGN), as suggested by  high-resolution cosmological simulations including cooling, star formation, supernova feedback, and AGN radio--mode feedback in galaxy clusters and groups (\citealt{puchwein08}, \citealt{bower08}, \citealt{short08}). 
Since supernova feedback appears to be insufficient to explain the L$_{X}$--T relation \citep{puchwein08}, feedback by AGN seems necessary.
According to this interpretation, gas can be removed from within R$_\mathrm{500}$ mainly as a consequence of the mechanical heating produced by a central AGN. The action of the AGN is larger in groups than in clusters simply because the potential well is  shallower in the former systems. In a forthcoming work we will quantify the feedback by AGN radio--mode for the COSMOS groups.
Another proposed mechanism capable of accounting for the "missing" gas is "filamentary heating" \citep{voit01}. Low entropy gas is consumed in star formation before the group formation, which eventually raises the entropy of the gas which becomes the ICM. The resulting higher entropy level inhibits the gas from falling towards the center of the potential well, which can explain the lack of gas in the central region of groups \citep{sun08}.

\section{Conclusions}

The baryon mass fraction is a parameter which can be constrained by the primordial light elements abundance set  by the nucleosynthesis at early epochs. It can be independently measured from observations of the CMB (e.g. WMAP) or of galaxy groups/clusters.  Different studies of the baryon mass fraction in nearby galaxy systems have reported values lower than the one from WMAP, the discrepancy being larger for groups than clusters. We investigate if this discrepancy persists when a sample of local clusters is supplemented by a large, unbiased sample of groups at $0.1\le z \le 1.0$.
Hereafter we list our conclusions.
\begin{enumerate}
\item The stellar mass fraction associated with galaxies is anticorrelated with the mass of the system: $f_\mathrm{500}^\mathrm{stars} \propto M_\mathrm{500}^{-0.37 \pm 0.04}$. This is consistent with previous results on local clusters. The validity of this result is now extended by one decade in total mass and to redshift 1. 
\item The previous relation holds after correcting the stellar mass fraction for a mass independent 11--22$\%$ contribution from the ICL as suggested by both observations and simulations. The slope of the $f_\mathrm{500}^\mathrm{stars}$--$M_\mathrm{500}$ relation is consistent with the constraint  set by the hierarchical paradigm of structure formation \citep{balogh08}. No significant evolution in the relation between f$_\mathrm{500}^\mathrm{stars}$ and M$_\mathrm{500}$ is observed. This supports the scenario in which massive clusters form mostly by merging of less massive groups and clusters, and observed groups in the redshift range 0--1 have formed the bulk of their stellar mass by z$\sim$1.0. 
\item Combining measured values of the stellar mass fraction with values of the gas mass fraction estimated from an average relation obtained for a local sample, f$_\mathrm{500}^\mathrm{stars+gas}$ increases by 25$\%$ from groups to clusters. After the introduction of appropriate corrections for gas depletion and ICL contribution, the total baryonic mass fraction at the groups regime still differs from the WMAP5 value at 3.3$\sigma$.
We interpret the origin of this discrepancy as a lack of gas (by 33$\%$), which can be produced either by feedback (supernovae and/or radio--mode AGN heating) or by  ''filamentary heating''.
\end{enumerate}
Our results provide useful constraints on simulations of the aforementioned processes. In particular the availability of a large unbiased sample of groups offers direct and stringent constraints on models rather than relying on extrapolation of the behaviour of the stellar fraction as a function of mass in the entire family of systems with $10^{13}<$M$_\mathrm{500}<10^{14}$M$_{\odot}$.
Future observations will increase both the statistics and the redshift sampling rate, so that a test and extension of our conclusions will be possible. 
 \acknowledgments
The authors thank the anonymous referee for her/his valuable comments, which led to a significant improvement of the paper. We acknowledge the contributions of the entire COSMOS collaboration; more informations on the COSMOS survey are available at \url{http://www.astr.caltech.edu/$\sim$cosmos}. This research was supported by the DFG Cluster of Excellence ÔOrigin and Structure of the UniverseÕ (\url{http://www.universe-cluster.de}). D.P. acknowledges support by the German \emph{Deut\-sches Zen\-trum f\"ur Luft- und Raum\-fahrt, DLR\/} project number 50~OR~0405.

\end{document}